\documentclass[amsfonts,showpacs,tightenlines,aps,12pt,floatfix]{revtex4}
\usepackage{bm}
\usepackage{epsfig}
\begin{document}


\title{Joint Extraction of $m_s$ and $V_{us}$ From Hadronic $\tau$ Decays}
\author{Kim Maltman}
\affiliation{Department of Mathematics and Statistics, York University, 
4700 Keele
St., Toronto, ON CANADA M3J 1P3}
\altaffiliation{CSSM, Univ. of Adelaide, Adelaide, SA 5005 AUSTRALIA}

\author{Carl E. Wolfe}
\affiliation{Department of Physics and Astronomy, York University, 
4700 Keele
St., Toronto, ON CANADA M3J 1P3}
\date{\today}

\begin{abstract}
We study the simultaneous determination of $m_s$ and $V_{us}$ 
from flavor-breaking hadronic $\tau$ decay sum rules
using weights designed to bring under better control problems associated 
with the slow convergence of the relevant $D=2$ OPE series.
Much improved stability and consistency is found 
as compared to the results of conventional analyses based on the 
``$(k,0)$ spectral weights''. Results for $m_s$ 
are in excellent agreement with those of recent strange scalar 
and strange pseudoscalar sum rule analyses, as well as
recent lattice analyses,
while those for $V_{us}$ agree within errors with the output from
recent lattice-based $\Gamma [K_{\mu 2}] /\Gamma [\pi_{\mu 2}]$ 
and $K_{\ell 3}$-based analyses.
Very significant error reductions are shown 
to be expected, especially for $V_{us}$, once the improved strange
spectral data from the B factory experiments becomes available.
\end{abstract}

\pacs{12.15Ff,12.15.Hh,13.35.Dx,11.55.Hx}

\maketitle

\section{\label{intro}Background}
Measurements of inclusive flavor $ij=ud,us$ vector (V) or axial vector (A) 
current induced hadronic $\tau$ decay distributions yield
kinematically weighted linear combinations of the spectral functions, 
$\rho^{(J)}_{V/A;ij}$, of the spin $J=0$ and $1$ parts,
$\Pi^{(J)}_{V/A;ij}$, of the relevant current-current correlators.
Explicitly, with 
$R_{V/A;ij}\, \equiv\, \Gamma [\tau^- \rightarrow \nu_\tau
\, {\rm hadrons}_{V/A;ij}\, (\gamma )]/ \Gamma [\tau^- \rightarrow 
\nu_\tau e^- {\bar \nu}_e (\gamma)]$~\cite{bnpetc},
\begin{equation}
R_{V/A;ij}\, =\, 12\pi^2\vert V_{ij}\vert^2 S_{EW}\,
\int^{m_\tau^2}_{th}\, {\frac{ds}{m_\tau^2}} \,
\left( 1-y_\tau\right)^2
\, \left[ \left( 1 + 2y_\tau\right)
\rho_{V/A;ij}^{(0+1)}(s) - 2y_\tau \rho_{V/A;ij}^{(0)}(s) \right]
\label{basictaudecay}\end{equation}
where $y_\tau =s/m_\tau^2$, $V_{ij}$ is the
flavor $ij$ CKM matrix element, $S_{EW}=1.0201\pm 0.0003$~\cite{erler02} 
is a short-distance electroweak correction, and the superscript
$(0+1)$ denotes the sum of $J=0$ and $J=1$ contributions.
Since the spectral function combinations in Eq.~(\ref{basictaudecay})
correspond to correlators with no kinematic singularities,
each term on the RHS can be rewritten
using the basic finite energy sum rule (FESR) relation, 
\begin{equation}
\int_{th}^{s_0}ds\, w(s) \rho (s)\, =\, {\frac{-1}{2\pi i}}\,
\oint_{\vert s\vert =s_0}ds\, w(s) \Pi (s)\ .
\label{fesrbasic}\end{equation}
Analogous FESR's, corresponding to spectral integrals, $R_{V/A;ij}^{(k,m)}$, 
obtained by rescaling the kinematic weights in
$R_{V/A;ij}$ by $(1-y_\tau )^ky_\tau^m$ before integration,
are referred to as the ``$(k,m)$ spectral weight sum rules''. 
Similar spectral integrals and FESR's can be 
constructed for $s_0<m_\tau^2$, for general non-spectral weights $w(s)$,
and for either of the correlator combinations $\Pi^{(0+1)}_{V/A;ij}(s)$ or
$s\Pi^{(0)}_{V/A;ij}(s)$. We denote such spectral integrals
generically by $R^w_{ij}(s_0)$, and refer to the purely $J=0$ 
contribution in ``inclusive'' FESR's (those having both 
$J=0+1$ and $J=0$ contributions) as ``longitudinal'', in what follows.

$V_{us}$ and/or $m_s$ are extracted using flavor-breaking 
differences, $\delta R^w(s_0)$, defined by
\begin{equation}
\delta R^w(s_0)\, =\, \left[R^w_{ud}(s_0)\, /\, \vert V_{ud}\vert^2\right]
\, -\, \left[R^w_{us}(s_0)\, /\, \vert V_{us}\vert^2\right]\ .
\label{tauvusbasicidea}\end{equation}
Since $\delta R^w(s_0)$ vanishes in the $SU(3)_F$ limit, its
OPE representation, $\delta R^w_{OPE}(s_0)$, begins with a
dimension $D=2$ contribution, proportional to $m_s^2$. 
Experimental values for $\delta R^w(s_0)$ over a range of $s_0$ and $w(s)$ 
allow $m_s$ and/or $V_{us}$ to be fitted, provided $s_0$ is large 
enough that insufficiently-well-known higher $D$ OPE contributions are  
small~\cite{taums,km00,longposviol,krmtau0204,pichetalvus,kmcwvus06,kmtau06}. 
As stressed in Refs.~\cite{pichetalvus,kmcwvus06,kmtau06}, the smallness
of $m_s$ (and hence of the flavor-breaking spectral integral differences)
presents a challenge for the determination of $m_s$ but a significant 
advantage for the determination of $V_{us}$. Explicitly, one has, from
Eq.~(\ref{tauvusbasicidea})~\cite{pichetalvus}, 
\begin{equation}
\vert V_{us}\vert \, =\, \sqrt{ R^w_{us}(s_0)/\Bigl( 
\left[ R^w_{ud}(s_0)/\vert V_{ud}\vert^2\right]
\, -\, \delta R^w_{OPE}(s_0)\Bigr)}\ .
\label{tauvussolution}\end{equation}
At scales $\sim 2-3\ {\rm GeV}^2$, and for weights used in the literature, 
the dominant $D=2$ term in $\delta R^w_{OPE}(s_0)$ is
{\it much} smaller than the separate $ud$, $us$ $D=0$ OPE contributions, 
and hence than the separate $ud$, $us$ spectral integrals (for physical 
$m_s$, typically at the few to several percent level). An uncertainty, 
$\Delta \left(\delta R^w_{OPE}(s_0)\right)$, in $\delta R^w_{OPE}(s_0)$ thus 
produces a fractional uncertainty in $\vert V_{us}\vert$,
$\simeq \Delta \left(\delta R^w_{OPE}(s_0)\right)/2\, R^w_{ud}(s_0)$, 
{\it much} smaller than that on $\delta R^w_{OPE}(s_0)$ itself. 
Moderate precision for $\delta R^w_{OPE}(s_0)$ thus suffices
for high precision on $\vert V_{us}\vert$, provided 
experimental errors can be brought under control.

In what follows, we perform a combined extraction of $m_s$ and 
$\vert V_{us}\vert$ based on existing spectral data. The $ud$ 
data~\cite{cleoud95,alephud98,opalud99} are already quite precise, 
but sizeable errors on the $us$ data~\cite{alephus99,cleous0305,opalus04}
limit currently achievable precision. We thus focus on better controlling
uncertainties on the theoretical (OPE) side of the analysis,
especially those associated with slower-than-previously-anticipated 
convergence of the relevant $D=2$ series~\cite{bck05}.
%

\section{\label{concerns}Technical issues in the hadronic $\tau$ 
decay approach}
%
The first major stumbling block is the very bad behavior of the 
integrated longtitudinal $D=2$ OPE series. Even at the maximum
scale $s_0=m_\tau^2$ allowed by kinematics, the series shows no
sign of converging~\cite{longconv}. Even worse, 
all truncation schemes employed in the literature, 
{\it with no exceptions}, badly violate constraints associated with 
spectral positivity~\cite{longposviol}. A large part of the very strong 
unphysical $k$-dependence seen in results for $m_s$ from inclusive $(k,0)$ 
spectral weight analyses was, in fact, associated with this spectral positivity
violation problem~\cite{longposviol}. Inclusive analyses
employing the longitudinal OPE representation are thus
untenable, and earlier results obtained using such analyses 
should be discarded in favor of those of more recent 
non-inclusive treatments~\cite{km00,pichetalvus,bck05}.

Fortunately, the severe problems of the longitudinal $D=2$ OPE
representation are easily handled phenomenologically, for a combination
of chiral and kinematic reasons. Apart from the $\pi$ and
$K$ pole contributions, longitudinal spectral contributions
vanish in the $SU(3)_F$ limit and are doubly-chirally suppressed
away from it. Since the values of the longitudinal kinematic weight
at the $K$ pole and in the region of the excited strange scalar and 
pseudoscalar (PS) resonances is essentially the same, this double chiral 
suppression is preserved in the ratio of non-pole to pole spectral 
integral contributions. The small residual non-pole $us$ PS and scalar 
contributions can, moreover, be well-constrained phenomenologically, 
the former via a sum rule analysis of the $us$ PS channel~\cite{km00,mkps}, 
the latter via $K\pi$-scattering-data-based dispersive 
analyses~\cite{cfnp,km00,jopss} (the most reliable being the
coupled-channel version discussed in Refs.~\cite{jopss}, which incorporates 
short-distance QCD and chiral constraints). With the very accurately known 
$\pi$ and $K$ pole contributions, these results allow a reliable bin-by-bin 
subtraction of longitudinal contributions to the experimental distribution, 
and hence a direct determination of the $(0+1)$ spectral function, from 
which FESR's not afflicted by the longitudinal $D=2$ OPE problem can be 
constructed. Note that agreement between the values of $m_s$ obtained from
scalar and PS sum rule analyses employing the resulting $us$ scalar and PS 
spectral ``models'' and those of recent $N_f=2+1$ lattice 
simulations~\cite{latticems} rules out significantly larger non-pole 
longitudinal spectral strength. The residual non-pole part of the longitudinal
subtraction is thus certainly small, and very well under control
at the level required for $(0+1)$ FESR analyses. 
We therefore focus, in what follows, on sum rules involving 
the flavor-breaking combination
$\Delta\Pi (s)\, \equiv\, \Pi_{V+A;ud}^{(0+1)}(s)\, -\, 
\Pi_{V+A;us}^{(0+1)}(s)$.

The second problem concerns the slow convergence of the $(0+1)$ $D=2$ OPE
series. For scales $s_0\sim 2-3\ {\rm GeV}^2$, $\left[\Delta\Pi\right]_{OPE}$
is dominated by its $D=2$ contribution~\cite{bck05},
\begin{equation}
\left[\Delta\Pi (Q^2)\right]^{OPE}_{D=2}\, =\, {\frac{3}{2\pi^2}}\,
{\frac{\bar{m}_s}{Q^2}} \left[ 1+2.333 \bar{a}+19.933 \bar{a}^2
+208.746 \bar{a}^3+(2378\pm 200)\bar{a}^4+\cdots \right]\ ,
\label{d2form}\end{equation}
where $\bar{a}=\alpha_s(Q^2)/\pi$ and $\bar{m}_s=m_s(Q^2)$, 
with $\alpha_s(Q^2)$ and $m_s(Q^2)$ the running coupling and strange quark 
mass in the $\overline{MS}$ scheme. The $O(\bar{a}^4)$ coefficient
has been estimated using approaches previously successful
in obtaining accurate predictions for the $O(\bar{a}^3)$ 
coefficient in Eq.~(\ref{d2form}) and $n_f$-dependent 
$O(\bar{a}^3m_q^2)$ coefficients of the electromagnetic current 
correlator in advance of the explicit calculations
of these values~\cite{bckzin04}.

Since, with 4-loop running, independent high-scale determinations
of $\alpha_s(M_Z)$~\cite{pdg06qcdreview} 
correspond to $\bar{a}(m_\tau^2)\simeq 0.10-0.11$, Eq.~(\ref{d2form}) 
shows that, at the spacelike point on $\vert s\vert =s_0$, the 
convergence of the $(0+1)$ $D=2$ OPE series is marginal at best,
even at the highest scales accessible in $\tau$ decay. While
$\vert \alpha_s(Q^2)\vert$ decreases as one moves along the contour 
away from the spacelike point, allowing the convergence of the integrated 
series to be improved through judicious weight choices,
this observation shows that, for weights not chosen specifically with 
this constraint in mind, one must expect to find very slow convergence 
of the integrated $D=2$ series. The $(k,0)$ spectral weights,
$w^{(k,0)}(y)=(1+2y)(1-y)^{k+2}$, with $y=s/s_0$, are very much
non-optimal in this regard, since 
$\vert 1-y\vert = 2\vert sin(\phi /2)\vert$ (where $\phi$ is the angular 
position measured counterclockwise from the timelike point), 
is peaked precisely in the spacelike direction. 
Slow convergence, deteriorating with increasing $k$, is thus expected 
for the integrated $D=2$ series of the $(k,0)$ spectral 
weights. The results of Table I of Ref.~\cite{bck05} and row 1,
Table I of Ref.~\cite{kmcwvus06} bear out this expectation.

In evaluating the integrated $(0+1)$ $D=2$ OPE contribution,
the size of the last term kept, 
the level of residual scale dependence, and the difference between the 
direct correlator and Adler function evaluations, both truncated at the 
same order, have all been used as measures of the truncation uncertainty.
The slow convergence of the integrated series, however, can make it hard to be
sufficiently conservative. For example, the quadrature sum
of the last term size plus residual scale dependence, used
previously in the literature to estimate the $O(\bar{a}^3)$ Adler function 
truncation uncertainty, yields a result $\sim 2.5$ times smaller than 
the actual difference between the $O(\bar{a}^3)$-truncated Adler function and 
$O(\bar{a}^4)$-truncated direct correlator results~\cite{kmcwvus06}.
Since the growth of $\alpha_s$ with decreasing scale makes
higher order terms relatively more important at lower scales,
premature truncation of a slowly-converging series typically shows up as
an unphysical $s_0$-dependence in extracted, nominally $s_0$-independent 
quantities. With polynomial weights, $w(y)=\sum_m c_my^m$, 
for which integrated $D=2N+2$ OPE contributions not suppressed by 
additional powers of $\alpha_s$ scale as $c_N/s_0^N$, 
such unphysical $s_0$-dependence can also result if unsuppressed 
higher $D$ contributions which might in principle be present 
are incorrectly assumed negligible and omitted from the analysis. 
Since, typically, not even rough estimates of
$D>6$ condensate combinations are available, such omission
is most dangerous for $w(y)$ having large values of
the coefficients $c_m$, with $m>2$, where such $D=2m+2>6$
contributions will be enhanced.
The $(2,0)$, $(3,0)$ and $(4,0)$ $J=0+1$ spectral weights,
$w^{(2,0)}(y)=1-2y-2y^2+8y^3-7y^4+2y^5$,
$w^{(3,0)}(y)=1-3y+10y^3-15y^4+9y^5-2y^6$,
and $w^{(4,0)}(y)=1-4y+3y^2+10y^3-25y^4+24y^5-11y^6+2y^7$,
are examples of weights having such large, potentially
dangerous higher order coefficients.

In view of the above discussion, $s_0$-stability tests 
are essential components of any FESR determination of 
$V_{us}$ and/or $m_s$. The existence of a stability window in $s_0$ for 
extracted quantities or, if not a stability window, then a window within 
which the observed instability is safely smaller than the estimated 
theoretical uncertainty, is crucial to establishing the reliability
of the theoretical error estimate.

\section{\label{results}The spectral and non-spectral weight analyses}
The OPE and spectral integral inputs used in our analysis are outlined below.
The fact that we restrict our attention to FESR's involving
the V+A spectral combination, to weights satisfying $w(s=s_0)=0$, 
and to scales $s_0>2\ {\rm GeV}^2$, all serve to strongly suppress 
possible residual OPE breakdown effects~\cite{krmtau0204,kmfesr,cdgmudvma}.

Integrals of the leading $D=2$ contribution to 
$\left[\Delta\Pi (Q^2)\right]_{OPE}$, given in Eq.~(\ref{d2form}),
are evaluated using two versions of the $O(\bar{a}^4)$-truncated CIPT 
prescription~\cite{ciptbasic}, one involving the RG-improved 
$D=2$ correlator contribution $\left[\Delta\Pi\right]^{OPE}_{D=2}$, 
the other the equivalent Adler function contribution,
$\left[\Delta D(Q^2)\right]^{OPE}_{D=2} \, =\, 
-Q^2 d\left[\Pi (Q^2)\right]^{OPE}_{D=2}/dQ^2$. The difference
between the two expressions, which would agree to all orders, 
but differ here by terms of $O(\bar{a}^{5})$ and higher,
is used as one measure of the truncation uncertainty. 
For $\bar{a}$ and $\bar{m}_s$ we employ
exact solutions corresponding to the 4-loop-truncated $\beta$ and 
$\gamma$ functions~\cite{betagamma4}, with initial condition 
$\alpha_s(m_\tau^2)=0.334\pm 0.022$. The remaining initial condition, 
$m_s(2\ {\rm GeV})$, is either taken as input or obtained as part of the fit. 

Expressions for the $D=4$ and $D=6$ contributions are given in 
Ref.~\cite{bnpetc}. The dominant $D=4$ term, proportional to 
$\langle m_s\bar{s}s \rangle$, 
is evaluated using ChPT quark mass ratios~\cite{leutwylerqm}, 
GMOR for the light quark condensate, and 
$r_c\equiv \langle m_\ell \bar{\ell}\ell\rangle /
\langle m_s\bar{s}s\rangle = 0.8\pm 0.2$. $D=6$ contributions are estimated
using the vacuum saturation approximation, and assigned an
uncertainty of $\pm 500\%$. Contributions with 
$D>6$ are assumed negligible, $s_0$-stability studies being
employed to test the self-consistency of this assumption.

For the spectral integrals we work with the ALEPH $ud$~\cite{alephud98} 
and $us$~\cite{alephus99} data, for which both data and covariance matrices are
publicly available. A small global renormalization of the $ud$
data is performed to reflect minor changes in the $e$, $\mu$ and total
strange branching fractions since the original ALEPH publication. 
Following the prescription of Ref.~\cite{chen01}, we also perform
mode-by-mode rescalings to reflect current (PDG06~\cite{pdg06})
values of the branching fractions of the various strange decay 
modes~\cite{thankstoshaomin}. 
Errors on the $K$ and $\pi$ 
pole contributions are reduced by using the more precise expectations 
based on  $\Gamma [\pi_{\mu 2}]$ and $\Gamma [K_{\mu 2}]$. 
With current data, $ud$ and $us$ spectral integral errors are at 
the $\sim 0.5\%$ and $\sim 3-4\%$ levels, respectively,
for the weights to be discussed below. BABAR and BELLE will drastically
reduce the size of the $us$ errors in the near future.

\subsection{\label{specwtsbad}The $(k,0)$ spectral weight analyses}
\begin{figure}
\unitlength1cm
\caption{$m_s$, $\vert V_{us}\vert$ joint fits using $s_0=m_\tau^2$
and various pairs of $(k,0)$ spectral weights}
\rotatebox{270}{\mbox{
\epsfig{figure=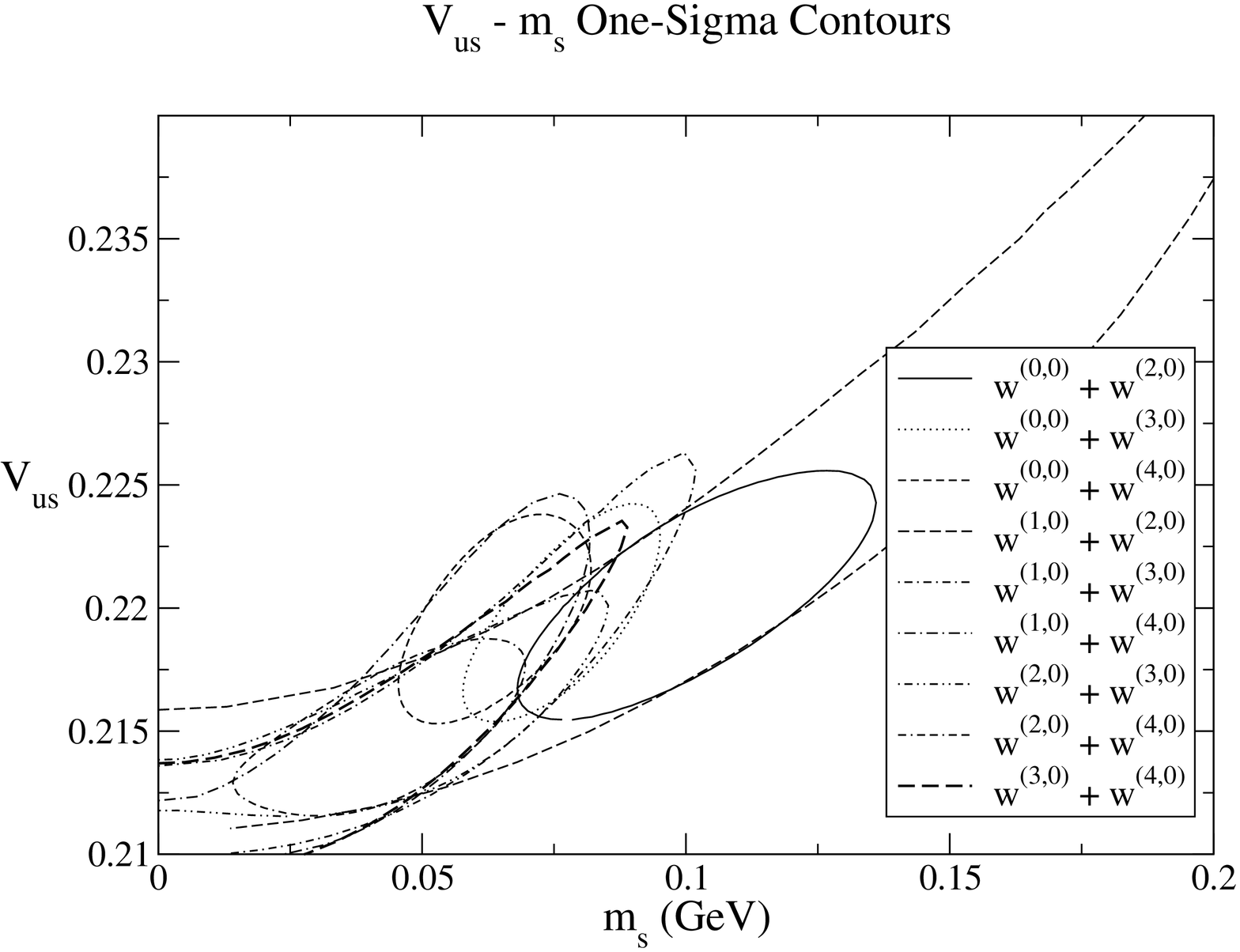,height=9.0cm,width=7.0cm}
}}
\label{figspecwtcontours}
\end{figure}

As noted above, slow convergence of the integrated $D=2$ OPE series
is both expected, and observed~\cite{bck05,kmcwvus06}, for the $(k,0)$ 
spectral weights. Further evidence that the OPE sides of the
$(k,0)$ sum rules are not under good control is
provided by Figure~\ref{figspecwtcontours}. The figure shows the
$1\sigma$ contours for joint fits of $m_s$ and $\vert V_{us}\vert$ 
to the $s_0=m_\tau^2$ experimental spectral integrals
for various pairs of $(k,0)$ spectral weights. 
It is clear that no sensible common fit region exists, and hence that
a reliable joint fit for $m_s$ and $\vert V_{us}\vert$ cannot be
obtained using the $(k,0)$ spectral weights considered previously
in the literature.

\begin{figure}
\unitlength1cm
\caption{$(0,0)$ spectral weight
OPE and spectral integrals vs. $s_0$ (GeV$^2$) for
various input $m_s(2\ {\rm GeV})$ and $\vert V_{us}\vert$ obtained
by matching the resulting OPE and spectral integrals at $s_0=m_\tau^2$}
\epsfig{figure=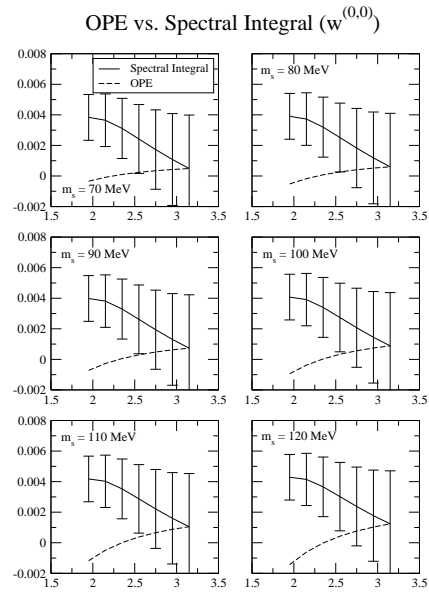,height=9.0cm,width=7.0cm}
\label{spec00msrange}
\end{figure}

Further consideration is warranted for
the $(0,0)$ analysis, which has been proposed in the literature 
as a particularly favorable one for the determination of 
$\vert V_{us}\vert$~\cite{pichetalvus}. Indeed, with $m_s$ from other 
sources as input, the $s_0=m_\tau^2$ version of this analysis, would,
if reliable, allow an improved determination of $\vert V_{us}\vert$
with only improved $us$ branching fractions as input, a feature
special to this weight and this $s_0$ value. 
Unfortunately, on the OPE side, the $D=2$ truncation uncertainty 
appears much larger than previously anticipated~\cite{kmcwvus06}, as
can be seen from (i) the rather poor agreement
between the truncated correlator and Adler function evaluations;
(ii) the strong deterioration in this situation with increasing
truncation order; and (iii) the very significant $s_0$-instability
in the output values of $\vert V_{us}\vert$, at least for the
PDG04 value of $m_s$ used as input in Ref.~\cite{kmcwvus06}. 
In Figure~\ref{spec00msrange} we demonstrate that this
$s_0$-instability is not an artifact of the particular $m_s$ employed 
in Ref.~\cite{kmcwvus06}. The figure shows the OPE and spectral integrals 
as a function of $s_0$ for the $(0,0)$ FESR, for a range of 
different fixed input $m_s$. The value of $\vert V_{us}\vert$ 
needed for this comparison was obtained by matching the OPE and 
spectral integral versions of $\delta R^{(0,0)}(m_\tau^2)$.
Note that very strong correlations exist amongst
the OPE integrals for different $s_0$ and, similarly, amongst the spectral
integrals for different $s_0$. Strongly discrepant $s_0$-dependences
for the OPE and spectral integrals, as seen in the figure 
{\it for the entire range of $m_s$ considered}, thus rule out the 
possibility that an acceptable $s_0$-stability in $\vert V_{us}\vert$ 
might be obtained from the $(0,0)$ analysis, in its current form,
for any input value of $m_s$.
Even restricting our attention to the rather narrow range of
$s_0$ values within $0.4\ {\rm GeV}^2$ of $m_\tau^2$,
the level of $s_0$-instability in $\vert V_{us}\vert$ is $>0.0020$,
more than a factor of $2$ larger than previous estimates of the
total theoretical uncertainty. 
In Ref.~\cite{kmcwvus06}, in order to obtain a more conservative estimate of
the truncation uncertainty, we employed {\it twice} the sum, 
in quadrature, of the last included term and the integrated
correlator-Adler function difference. The resulting estimated
truncation uncertainty (total theoretical uncertainty) in 
$\vert V_{us}\vert$ was then $\pm 0.0020$ ($\pm 0.0022$) for $s_0=m_\tau^2$,
compatible at least with the observed $s_0$-instability in the limited
$s_0$ region noted above.
While this estimate is considerably more conservative than those 
employed previously in the literature, we argue that, given the
observed level of $s_0$-instability, less cautious assessments
can not be justified. Since the $D=2$ truncation uncertainty is
unlikely to be reduced, we conclude that, unfortunately,
theoretical uncertainties preclude a $(0,0)$ spectral
weight determination of $\vert V_{us}\vert$ at the sub-$\pm 0.0020$ level.
Fortunately, alternate weights with significantly 
improved $D=2$ convergence behavior exist, allowing one to take advantage of
the general approach proposed in Refs.~\cite{pichetalvus}. 

\subsection{\label{nonspecwtsgood}The non-spectral-weight analyses}
In Ref.~\cite{km00}, three non-spectral polynomial weights,
$w_{10}(y)$, $\hat{w}_{10}(y)$, and $w_{20}(y)$, designed specifically to
produce improved integrated $D=2$ convergence, were constructed. 
The significantly improved convergence is displayed explicitly
in Ref.~\cite{km00}, and Table 1 of Ref.~\cite{kmcwvus06}.
The weights were also designed to (i) keep
higher order coefficients which might enhance 
$D>6$ contributions small, and, (ii) strongly 
suppress spectral integral contributions from the region
above $s\sim 1\ {\rm GeV}^2$, where $us$ errors are large.
Here we consider also a fourth weight, $w_8(y)$, with lower degree,
but less strong suppression of the high-$s$ part of the $us$ spectrum.
While the latter feature leads to larger $us$ spectral integral errors
with present data, the weight will be useful for future analyses 
based on data with smaller $us$ spectral errors. Having an additional
non-spectral weight also allows us to investigate more fully the issue of
the mutual consistency of analyses associated with different
improved-convergence non-spectral weight choices.

\begin{figure}
\unitlength1cm
\caption{$w_{20}$ and $w_{10}$ OPE and spectral integrals 
(GeV$^2$) vs. $s_0$ for various input $m_s(2\ {\rm GeV})$ and 
$\vert V_{us}\vert$ obtained by matching the resulting OPE and spectral 
integrals at $s_0=m_\tau^2$}

\begin{minipage}[t]{7.2cm}
\begin{picture}(7.1,11.1)
\epsfig{figure=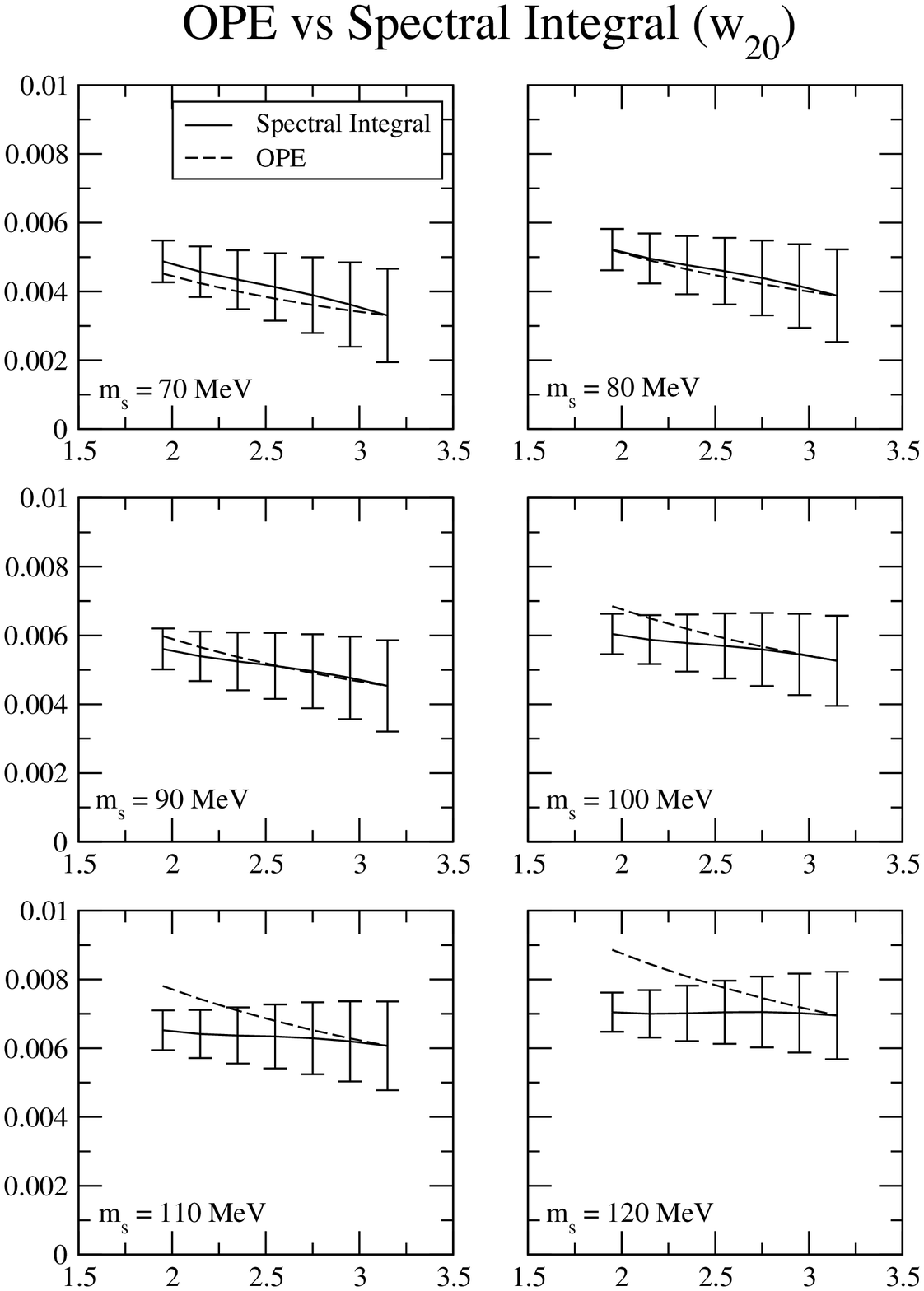,height=11.0cm,width=7.0cm}
\end{picture}
\end{minipage}
\hfill
\begin{minipage}[t]{7.2cm}
\begin{picture}(7.1,11.1)
\epsfig{figure=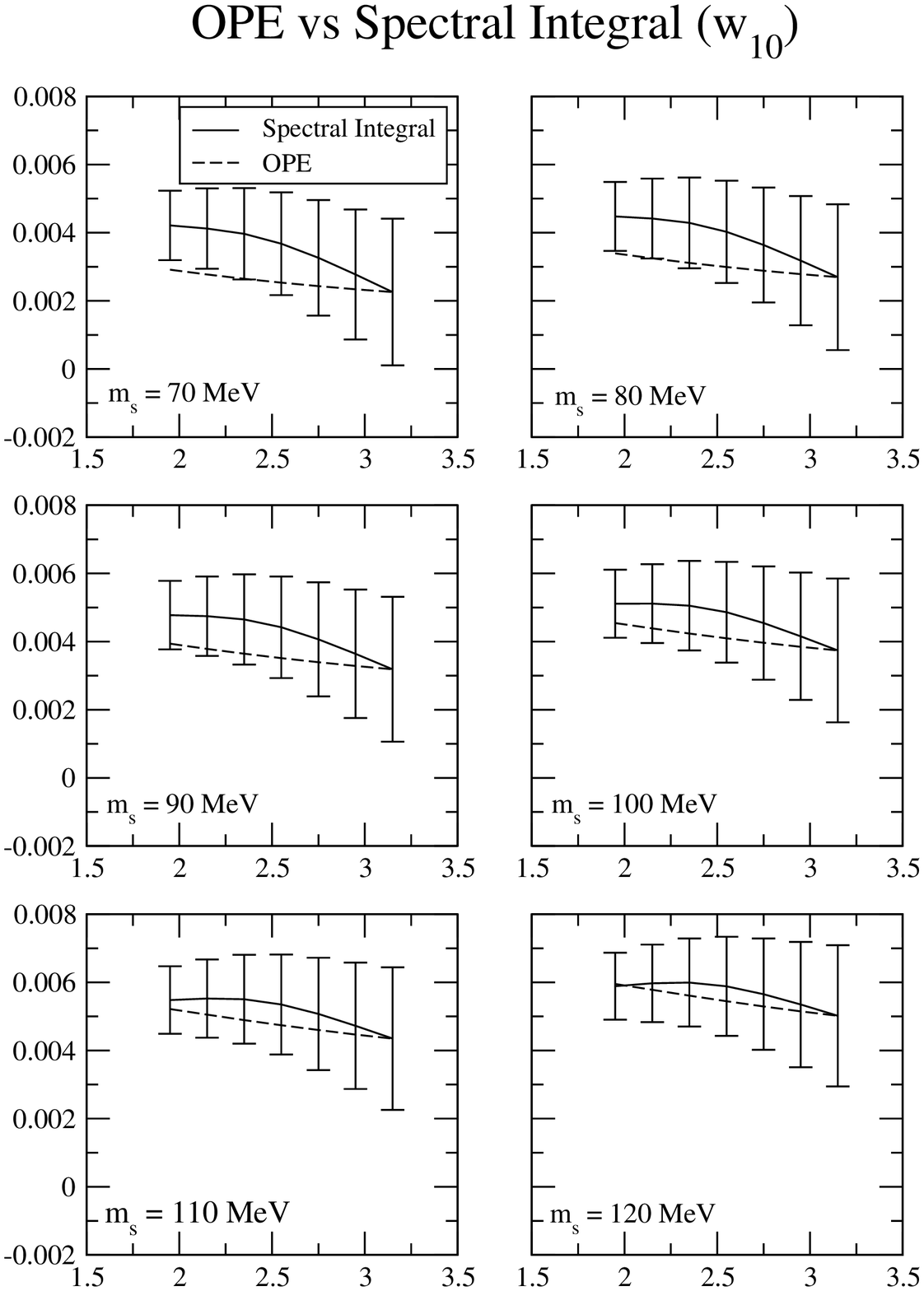,height=11.0cm,width=7.0cm}
\end{picture}
\end{minipage}
\label{nonspecwtmsrangeplots}
\end{figure}

In Figure~\ref{nonspecwtmsrangeplots} we display the matches between the OPE 
and spectral integral differences for $w_{20}$ and $w_{10}$,
for a range of input $m_s(2\ {\rm GeV})$. The spectral integral
differences are computed as described above for the analogous $(0,0)$ 
spectral weight case, shown in Figure~\ref{spec00msrange}. 
The $\hat{w}_{10}$ and $w_8$ matches are intermediate in quality,
and hence not shown separately. The contrast with the $(0,0)$ 
spectral weight case is immediately evident. 

\begin{figure}
\unitlength1cm
\caption{2-, 3- and 4-fold, $s_0=m_\tau^2$, joint $m_s$, 
$\vert V_{us}\vert$ non-spectral weight fits}
\rotatebox{270}{\mbox{
\epsfig{figure=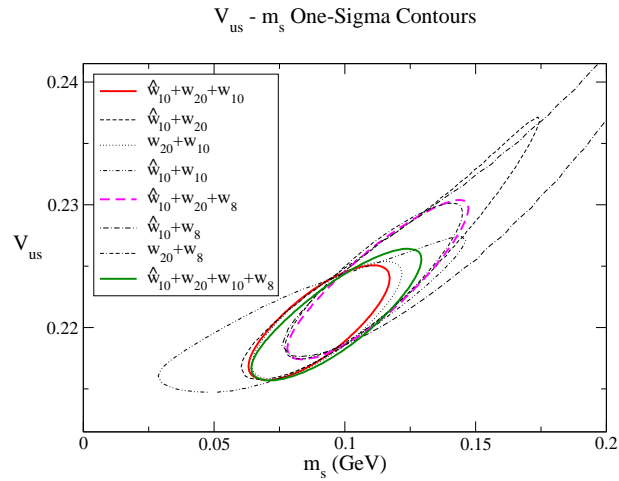,height=9.0cm,width=7.0cm}
}}
\label{nonspecwtcontours}
\end{figure}

In Figure~\ref{nonspecwtcontours} we display the joint fit,
$s_0=m_\tau^2$, $1\sigma$ contours, analogous to those of 
Figure~\ref{figspecwtcontours}, for various pairs of
the non-spectral weights. Also shown are the $1\sigma$
contours for two 3-fold and the full 4-fold fits.
A good common fit region for $m_s$ and
$\vert V_{us}\vert$ obviously exists,
in sharp contrast to the situation for the $(k,0)$ spectral
weights. 

We first consider results for $\vert V_{us}\vert$
obtained, as in previous studies, using
single weight analyses with $s_0=m_\tau^2$ and external $m_s$ input. With 
the average of strange scalar and PS sum rule and
$N_f=2+1$ lattice results, $m_s(2\ {\rm GeV})=94\pm 6$ MeV,
advocated in the last of Refs.~\cite{jopss}, one finds
$\vert V_{us}\vert =0.2209\pm 0.0029_{exp}\pm 0.0017_{th}$ for $w_{20}$, 
$0.2210\pm 0.0030_{exp}\pm 0.0010_{th}$ for $\hat{w}_{10}$, 
$0.2206\pm 0.0032_{exp}\pm 0.0007_{th}$ for $w_{10}$, 
and $0.2218\pm 0.0037_{exp}\pm 0.0009_{th}$ for $w_{8}$,
with large experimental errors, dominated by those on the 
$us$ distribution. The results are consistent, but cannot 
be directly averaged due to strong correlations. 

A cautious approach, given the instrinsically slow convergence of 
the $(0+1)$ $D=2$ series, is to ignore external information on
$m_s$, perform a combined fit for $m_s$ and $\vert V_{us}\vert$,
and verify that the results for $m_s$ are consistent with what
is known from other sources. The results for the 3-fold, $s_0=m_\tau^2$,
$w_{20}$, $\hat{w}_{10}$, $w_{10}$ fit, including theoretical
errors on the same footing as experimental ones in the
minimization process, are $m_s(2\ {\rm GeV})=89\pm 26$ MeV
and $\vert V_{us}\vert =0.2202\pm 0.0046$. The analogous full
4-fold non-spectral weight fit yields $m_s(2\ {\rm GeV})=96\pm 31$ MeV
and $\vert V_{us}\vert =0.2208\pm 0.0052$. The larger errors in
the second case result from the less strong suppression of the
high-$s$ $us$ data region by $w_8$. 

Since the combined joint fit results for
$m_s$ are in excellent agreement with the external average noted
above, it makes sense now to perform a combined $s_0=m_\tau^2$
fit for $\vert V_{us}\vert$ with $m_s(2\ {\rm GeV})=94\pm 6$ MeV
as input. The lower-error 3-fold fit (without $w_8$) yields
\begin{equation}
\vert V_{us}\vert =0.2209\pm 0.0031\ ,
\label{vusmainresult}\end{equation}
which we take as our main result. Note that, had the fit been performed
without including the uncertainty on the central value for 
$m_s$, the output central value for $\vert V_{us}\vert$
would have been shifted up to $0.2221$. 
The result of Eq.~(\ref{vusmainresult}), though lower in central
value, is compatible, within
mutual errors, with both the recent $K_{\ell e}$ determination,
$\vert V_{us}\vert = 0.2249\pm 0.0019$~\cite{antonelli06}
(based on the conventional Leutwyler-Roos estimate for
$f_+(0)$)) and the $\Gamma [K_{\mu 2}]/\Gamma [\pi_{\mu 2}]$
determination, $\vert V_{us}\vert = 0.2223^{+0.0026}_{0.0013}$
(based on the updated 2006 MILC evaluation of $f_K/f_\pi$~\cite{latticems}).
We comment further below on the status/reliability of the 
current hadronic-$\tau$-decay-based analysis.

\section{\label{conclusions}Discussion, conclusions and prospects}
We have shown that it is possible to construct non-spectral
weights which improve the convergence behavior of the integrated
$D=2$ OPE series and allow a self-consistent joint fit for 
$m_s$ and $\vert V_{us}\vert$. While the results reported above
for $\vert V_{us}\vert$ are compatible within errors with those
from other sources, one should bear in mind that the errors
on the present hadronic $\tau$ decay data are sufficiently
large that non-trivial shifts in the central value are
likely when the B-factory data becomes available. To see why
this is the case, note that the branching fractions
for observed strange decay modes in the current data go down
only to the $\sim 3\times 10^{-4}$ level. Missing modes with lower branching 
fractions could have a non-trivial impact on the extracted value of 
$\vert V_{us}\vert$. Missing a strange mode with branching fraction 
$1\times 10^{-4}$, for example, would lower $\vert V_{us}\vert$ 
by $\sim 0.0004$. At the
desired level of precision, one thus needs to detect all strange
modes with branching fractions at the few-to-several-$10^{-5}$ level.
Such determinations are certainly feasible at the B-factory experiments
where preliminary results for branching fractions at this level
(with errors at the few-to-several-$10^{-6}$ level) have already
been reported~\cite{babarbelleprelim}. 

Regarding the goal of improved future precision, one
should note that the $\sim 1/2\%$ uncertainty
on the $ud$ spectral integrals, though negligible in comparison
to the $us$ errors at present, contributes $\sim 0.0005$ to 
the uncertainty on $\vert V_{us}\vert$, assuming the current
assessment of the non-strange spectral errors is correct.
One should also bear in mind here the disagreement
between the isovector vector spectral contributions implied
by isospin-breaking electroproduction data and those measured
directly in $\tau$ decay experiments~\cite{daviertau06}.
For example, were the electroproduction results for
$\pi\pi$ (whose implied $\tau\rightarrow\pi\pi\nu_\tau$ 
branching fraction is $4.5\sigma$ below that measured directly 
in $\tau$ decay) to be correct, the value of $\vert V_{us}\vert$
obtained from the $\tau$ decay analysis above would be
raised by $\sim 0.0018$. Even the slightly lower central
value of the preliminary BELLE $\tau\rightarrow\pi\pi\nu_\tau$ 
measurement (which is compatible, within its somewhat larger
systematic error, with the earlier $\tau$
world average value) would produce an increase of
$\sim 0.0006$ in $\vert V_{us}\vert$. Resolving the discrepancy between
the $\pi\pi$ electroproduction and $\tau$ decay results
is thus important not just for clarifying expectations for
the hadronic contribution to $(g-2)_\mu$ in the Standard Model, 
but also for finalizing future $\tau$ decay determinations of 
$\vert V_{us}\vert$.

We finally comment that, re-running the analysis with 
the central values of the $us$ spectral distribution
unchanged but with errors reduced by a factor of $5$ 
(covariances by a factor of 25), and 
assuming $m_s(2\ {\rm GeV})$ to be known to $\pm 5\ {\rm MeV}$, 
a combined total fit error below $0.0010$ is achieved. Such an
improvement in the $us$ spectral distribution errors should be
well within reach of the final B-factory analyses~\cite{simonprivate}.
Further improvement on the theoretical component of the total
error is almost certainly also possible through the use of new weights with
significantly less strong suppression of the high-$s$ region. Such
weights, however, will become useful only once the errors in the $us$
spectrum above the $K^*$ have been significantly reduced. 
The necessity of self-consistency and stability checks also means
that an interaction between theorists and experimentalists
will certainly be required for successful development of such weights.

\begin{acknowledgments}
Ongoing support from the Natural Sciences and Engineering
Research Council of Canada is gratefuly acknowledged.
\end{acknowledgments}

\vfill\eject

\begin{thebibliography}{99}
\bibitem{daviertau06}See, e.g., M. Davier's review of the situation
at Tau'06 (tau06.sns.it).
\end{thebibliography}
\end{document}